\documentclass[letters,useAMS,usenatbib]{mn2e}
\usepackage{graphicx}

\def\gtrsim{\mathrel{\hbox{\rlap{\hbox{\lower4pt\hbox{$\sim$}}}\hbox{$>$}}}}

\begin{document}
\title[Strong-lensing analysis of Abell 370]{
Abell 370 revisited: refurbished Hubble imaging of the first strong lensing cluster
}
\author[Richard et al.]{
 \parbox[h]{\textwidth}{
J. Richard$^{1}$\thanks{Marie-Curie fellow. E-mail:
johan.richard@durham.ac.uk}, J.-P. Kneib$^{2}$,
M. Limousin$^{2,3}$,
A. Edge$^{1}$,
E. Jullo$^{4}$
}
\vspace{6pt}\\
$^{1}$Institute for Computational Cosmology, Department of Physics, Durham University,
South Road, Durham, DH1 3LE, UK \\
$^{2}$Laboratoire d'Astrophysique de Marseille, CNRS- Universit\'e Aix-Marseille, 38 rue Fr\'ed\'eric Joliot-Curie, 13388 Marseille Cedex 13, France\\
$^{3}$Dark Cosmology Centre, Niels Bohr Institute, University of Copenhagen,
         Juliane Maries Vej 30, 2100 Copenhagen, Denmark\\
$^{4}$Jet Propulsion Laboratory, Caltech, MS 169-327, Oak Grove Dr, Pasadena CA 91109, USA\\
}

\date{Accepted 2009 November 27. Received 2009 November 27; in original form 2009 October 29.}

\pagerange{\pageref{firstpage}--\pageref{lastpage}} \pubyear{2002}

\maketitle

\label{firstpage}

\begin{abstract}

We present a strong lensing analysis of the galaxy cluster Abell 370 (z=0.375) based on the recent multicolor ACS images obtained as part of the Early Release Observation (ERO) that followed the Hubble Service Mission \#4. 
Back in 1987, the giant gravitational arc ($z=0.725$) in Abell 370 was one of the first pieces of evidence that massive clusters are dense enough to act as strong gravitational lenses. The new observations reveal in detail its disklike morphology, and we show that it can be interpreted as a complex five-image configuration, with a total magnification factor of $32\pm4$. Moreover, the high resolution multicolor information allowed us to identify 10 multiply imaged background galaxies.

 We derive a mean Einstein radius of $\theta_{E}=39\pm2$\arcsec\ for a source redshift at $z=2$, corresponding to a mass of M$(<\theta_{E}) = 2.82\pm0.15\times 10^{14}$ M$_{\odot}$ and M$(<250\ {\rm kpc})=3.8\pm0.2\times 10^{14}$ M$_{\odot}$, in good agreement with Subaru weak-lensing measurements. 
The typical mass model error is smaller than 5\%, a factor 3 of improvement compared to the previous lensing analysis.
Abell 370 mass distribution is confirmed to be bi-modal with very small offset between the dark matter, the X-ray gas and the stellar mass. Combining this information with the velocity distribution reveals that Abell 370 is likely the merging of two equally massive clusters along the line of sight, explaining the very high mass density necessary to efficiently produce strong lensing.

These new observations stress the importance of multicolor imaging for the identification of multiple images which is key to determining an accurate mass model. 
The very large Einstein radius makes Abell 370 one of the best clusters to search for high redshift galaxies through strong magnification in the central region.

\end{abstract}

\begin{keywords}
Gravitational lensing - 
Galaxies: clusters: general - 
Galaxies: clusters: individual (A370)
\end{keywords}

\section{Introduction}

The discovery of the ``giant luminous arcs'' in rich clusters of galaxies in the mid-80's \citep{Lynds86,Soucail87} has opened a new window of research in cosmology: gravitational lensing. This simple geometric tool allows to map out dark matter in the Universe on various angular scales. The strong lensing regime occurs in the densest part of galaxies and massive clusters. 
When a background source is straddling one or more caustic lines, giant luminous arcs can be produced. 
The identification of multiple images has remained difficult untill deep multi-color space-based imaging has become routinely possible. This has been effectively the case with the installation of the Advanced Camera for Surveys (ACS) aboard  {\it Hubble} in March 2002. ACS observations of lensing clusters have lead to many discoveries of tens of multiple images (A1689: \citealt{Broadhurst05}, \citealt{Limousin07}; A1703: \citealt{Limousin08}, \citealt{Richard09}; RXJ1347: \citealt{Bradac08}; the bullet cluster: \citealt{Bradac07}; A2218: \citealt{Ardis}; Cl0024+1654: \citealt{Zitrin09a}; MS1358: \citet{Swinbank09}; MACS clusters: \citealt{Zitrin09b}, Smith et al 2009, Limousin et al 2010). These discoveries have foster the development of new mass modeling techniques (e.g. \citealt{Jullo}, \citealt{Coe}, \citealt{Jullo09}) that take advantage of the numerous constraints coming from these many multiple images.
The accuracy of the best mass model is now approaching the percent level allowing: (1) the use of massive clusters lenses as probes of cosmography -- hence putting strong geometrical constraints on the cosmological model (\citealt{Golse2002}, \citealt{Soucail04}, Jullo \& Kneib 2009), (2) recovering the intrinsic shape of lensed galaxies \citep{Swinbank09}, and (3) accurate correction of the magnification and dilution effect of massive clusters to constrain the luminosity function of distant galaxies (e.g. \citealt{Richard08}, \citealt{Bouwens09}).

The ACS failure in January 2007 has stopped the observation and detailed analysis of further massive clusters. The recent recovery of the ACS camera last May 2009 now brings new opportunities to further investigate massive cluster lenses.
We present in this letter a strong lensing analysis of the multicolor cluster observations obtained on Abell 370 as part of the Early Release Observation validating the performance of the repaired ACS instrument (Sect. \ref{data}).  The identification of 10 multiply imaged background galaxies (hereafter called systems of multiple images) totaling 32 images and the lens modeling are presented in Sect. \ref{sl}. The results on the nature of the giant arc and the mass distribution and substructure are presented in Sect \ref{results}.

Throughout the paper, we use  magnitudes quoted in the AB system, and a standard $\Lambda$-CDM model with $\Omega_m=0.3$, 
$\Omega_\Lambda=0.7$, and $H_0=70$ km s$^{-1}$ Mpc $^{-1}$, whenever necessary. 1\arcsec\ on the sky is equivalent to a physical distance of 5.16 kpc at the redshift $z=0.375$ of the galaxy cluster.

\section{Observations and data reduction}
\label{data}

The HST/ACS observations were obtained in July 2009 (PID 11507, PI: Noll), for a total of 6780/2040/3840 sec in the F475W/F625W/F814W bands, 
respectively. We make use of the images provided by STScI, reduced using the package multidrizzle,
and we measure AB depths (3 $\sigma$) of 28.39/27.41/27.84 for a point source.
The astrometry of these images was checked using the USNO optical catalogs.

\begin{figure*}
\centerline{\mbox{\includegraphics[width=0.9\textwidth,angle=0]{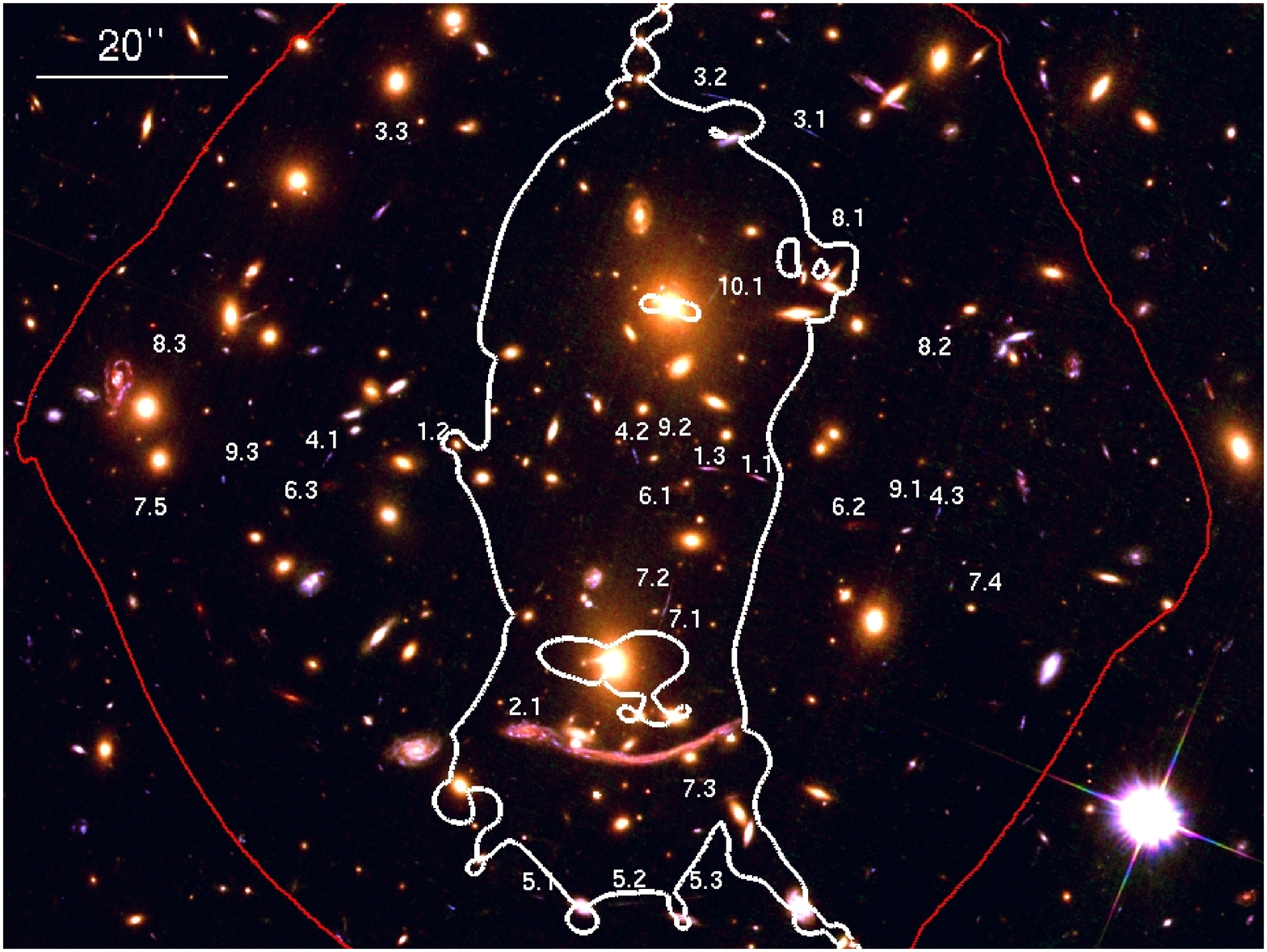}}}
\centerline{\mbox{\includegraphics[width=0.60\textwidth,angle=0]{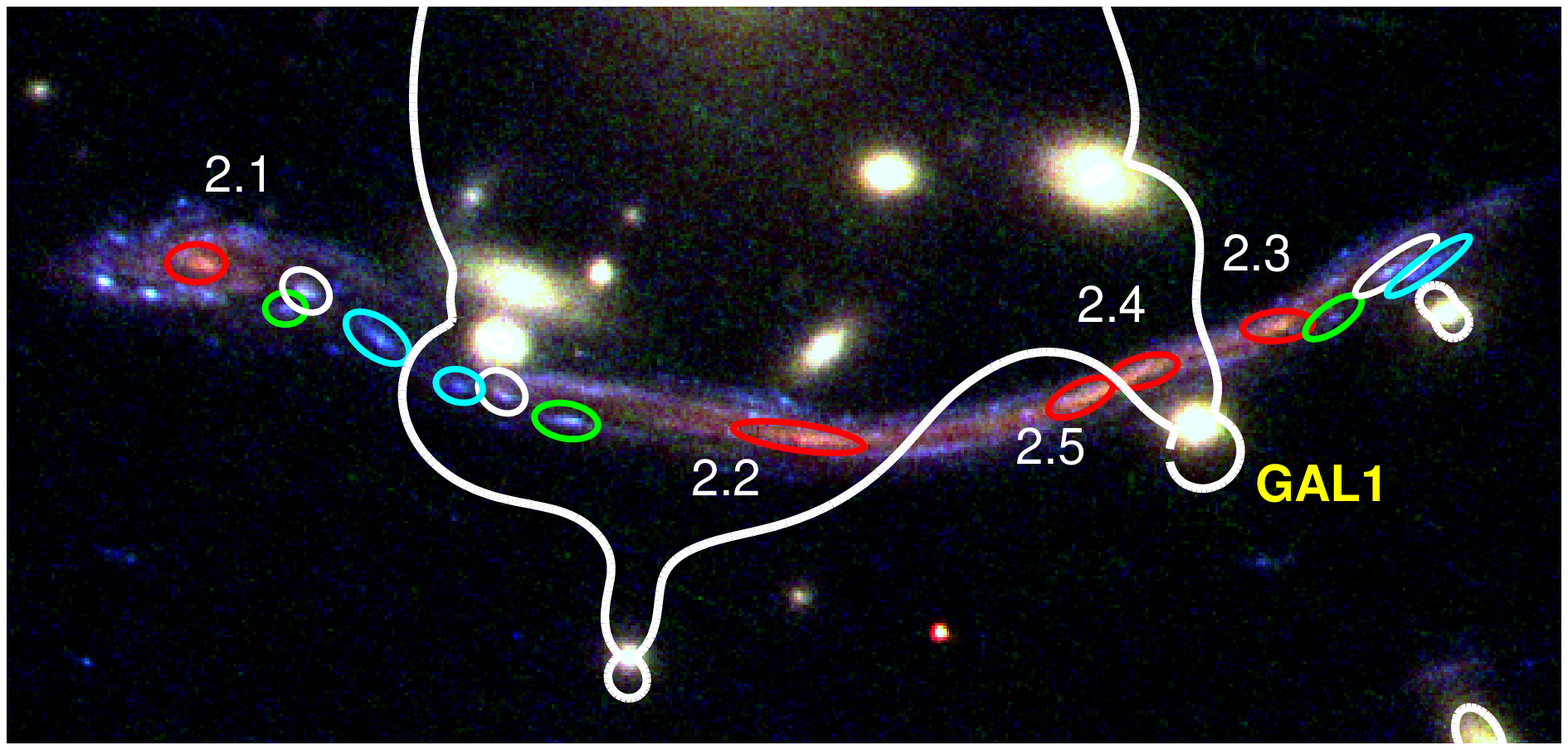}\hspace{0.5mm}
\includegraphics[width=0.29\textwidth,angle=0]{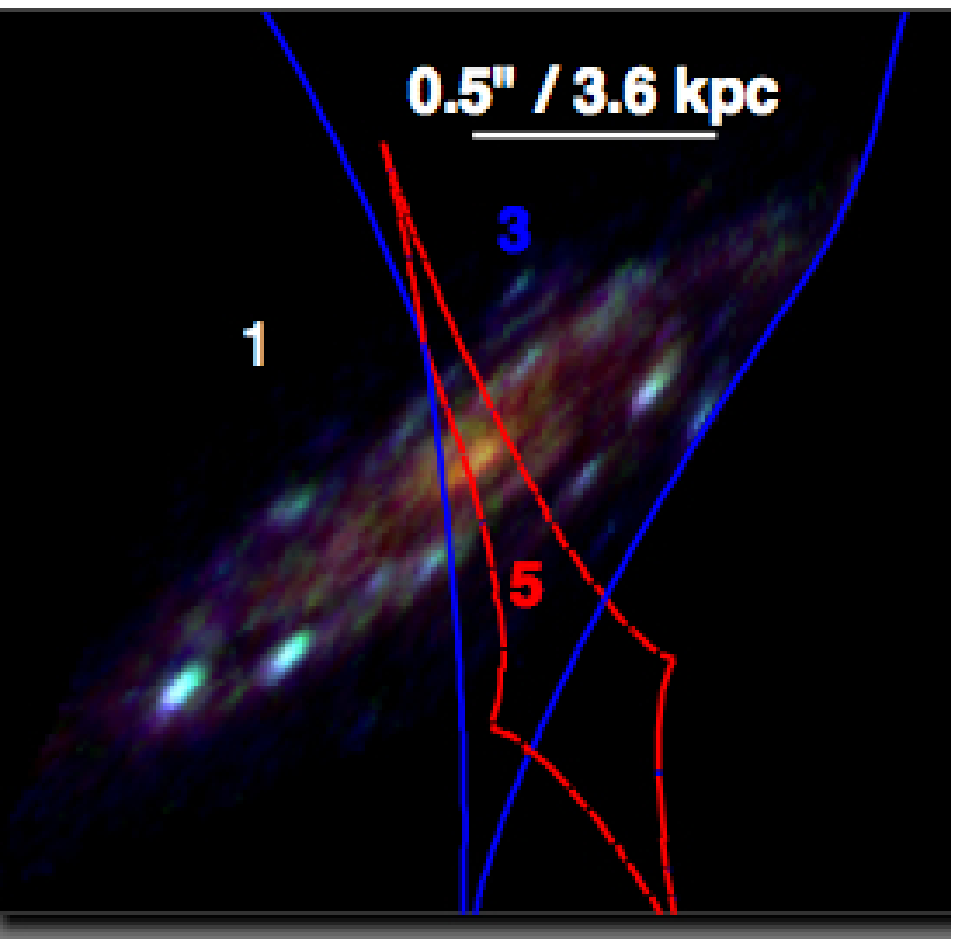}
}}
\caption{\label{mulfig}(Top panel) ACS F475W/F625W/F814W color image showing the location of the identified multiple systems, along with 
the critical line at z=1.2 (redshift for the majority of the mutiple images). The red line delimits the region of multiple images for very high 
redshift sources (here assuming $z=6$).
(Bottom panels) System 2: Location of matched regions (identified with colored ellipses) in each part of the giant arc. The white line shows the critical curve at $z=0.725$. The right panel shows the  source plane reconstruction for images 2.1. Overlayed are the 
caustic curve originated from the cluster-scale clump (blue) and the galaxy GAL1 (red), delineating three regions of image 
multiplicity.
}
\end{figure*}

We performed a visual inspection of the images to search for multiply lensed background galaxies. 
We made use of previous strong-lensing work by \citet{Kneib93}, \citet{Smail96} and \citet{Bezecourt1} to guide our multiple image
identification. Compared to the single WFPC2 observations in F625W, the new multicolor ACS data is deeper 
and has higher resolution. We identify 10 secure systems of images, including 6 new confirmed systems compared to previous lensing analyses(e.g. \citealt{Abdelsalam}), which we present in Fig. \ref{mulfig} and Table \ref{mult}. We confirm that all these images are genuine by reproducing their positions and shape using the 
constructed lens model (Sect. \ref{sl}) and by checking that their ACS colors agree within the 1$\sigma$ error bar (Table \ref{mult}). 
We use the notation $\alpha .\beta$ to design image $\beta$ of system $\alpha$. Two of these systems (1 and 2) have spectroscopic redshifts published by \citet{Soucail} and \citet{Kneib93b}.
 Seven systems (all except 2, 7 and 10) are tangential systems of 3 images, 
while systems 7 and 10 are two radial arcs. We identify 5 images in total for system 7, but system 10 has a much lower surface 
brightness and we can not identify its counter-images securely. System 2 is the giant arc first identified by \citet{Soucail87}, 
and is a lensed image of a spiral galaxy. Thanks to the new multicolor images we identify 5 images of the central red bulge, 
while individual star-forming regions are recognized as triple images with blue colors (Fig. \ref{mulfig}). 

\begin{table*}
\caption{\label{mult}Properties of the multiple imaged systems. Redshift values quoted with brackets are predictions from 
the lensing model. } 

\begin{tabular}{lrrrrllllll}
Multiple & $\alpha$ & $\delta$ & F814W$-$F475W & F814W$-$F625W & F814W & $z$ \\
\hline
1.1 & 39.966857 & -1.5769063 & $-1.08\pm0.13$ & $-0.66\pm0.10$ & $23.04\pm0.06$ & 0.806 \\
1.2 & 39.976088 & -1.5760316 & $-0.93\pm0.17$ & $-0.66\pm0.14$ & $23.88\pm0.08$ & 0.806 \\
1.3 & 39.968466 & -1.5766097 & $-1.11\pm0.13$ & $-0.77\pm0.10$ & $22.96\pm0.06$ & \\ 
\hline
2.1 & 39.973639 & -1.5842121 & $-2.61\pm0.24$ & $-1.13\pm0.08$ & $21.81\pm0.05$ & 0.725 \\
2.2 & 39.970772 & -1.5850722 & $-2.54\pm0.17$ & $-1.04\pm0.06$ & $21.48\pm0.05$ & 0.725 \\
2.3 & 39.968546 & -1.5845042 & $-2.65\pm0.15$ & $-1.07\pm0.06$ & $20.88\pm0.05$ & 0.725 \\
2.4 & 39.969185 & -1.5847126 & $-2.65\pm0.15$ & $-1.07\pm0.06$ & $20.88\pm0.05$ & 0.725 \\
2.5 & 39.969421 & -1.5848306 & $-2.20\pm0.09$ & $-0.55\pm0.06$ & $21.88\pm0.05$ & 0.725 \\
\hline
3.1 & 39.965457 & -1.5668639 & $ 0.31\pm0.23$ & $ 0.39\pm0.22$ & $24.08\pm0.18$ & [1.59$\pm$0.07] \\
3.2 & 39.968333 & -1.5658229 & $ 0.19\pm0.25$ & $ 0.52\pm0.21$ & $24.05\pm0.18$ & \\ 
3.3 & 39.977084 & -1.5671820 & $ 0.00\pm0.37$ & $ 0.28\pm0.35$ & $25.16\pm0.25$ & \\ 
\hline
4.1 & 39.979437 & -1.5763194 & $-0.03\pm0.18$ & $ 0.05\pm0.18$ & $24.37\pm0.13$ & [1.26$\pm$0.03] \\
4.2 & 39.970545 & -1.5762639 & $ 0.02\pm0.21$ & $-0.01\pm0.22$ & $24.38\pm0.15$ & \\ 
4.3 & 39.961736 & -1.5779306 & $-0.08\pm0.27$ & $-0.15\pm0.28$ & $24.81\pm0.18$ & \\ 
\hline
5.1 & 39.973275 & -1.5890694 & $-0.30\pm0.36$ & $-0.41\pm0.38$ & $24.69\pm0.21$ & [1.28$\pm$0.03] \\
5.2 & 39.970954 & -1.5892361 & $-0.78\pm0.55$ & $-0.45\pm0.19$ & $23.53\pm0.16$ & \\ 
5.3 & 39.968620 & -1.5890278 & $-0.46\pm0.30$ & $-0.71\pm0.36$ & $23.93\pm0.16$ & \\
\hline
6.1 & 39.969266 & -1.5771944 & $-2.32\pm0.72$ & $-1.39\pm0.31$ & $23.17\pm0.09$ & [1.06$\pm$0.02] \\
6.2 & 39.964320 & -1.5782500 & $-2.75\pm0.87$ & $-1.56\pm0.47$ & $22.66\pm0.08$ & \\ 
6.3 & 39.979437 & -1.5771388 & $-2.60\pm0.44$ & $-1.69\pm0.33$ & $22.46\pm0.08$ & \\ 
\hline
7.1 & 39.969579 & -1.5804097 & $-0.56\pm0.09$ & $-0.18\pm0.08$ & $23.03\pm0.06$ & [1.78$\pm$0.25] \\
7.2 & 39.969673 & -1.5807406 & $-0.46\pm0.38$ & $-0.16\pm0.31$ & $24.97\pm0.20$ & \\ 
7.3 & 39.968606 & -1.5856111 & $-0.89\pm0.45$ & $-0.23\pm0.30$ & $24.85\pm0.20$ & \\ 
7.4 & 39.961367 & -1.5800139 & $-0.64\pm0.30$ & $ 0.04\pm0.18$ & $24.54\pm0.10$ & \\ 
7.5 & 39.983602 & -1.5779652 & $-0.48\pm0.52$ & $-0.23\pm0.43$ & $25.54\pm0.26$ & \\ 
\hline
8.1 & 39.964292 & -1.5697917 & $-0.20\pm0.54$ & $ 0.36\pm0.42$ & $25.91\pm0.33$ & [2.41$\pm$0.22] \\
8.2 & 39.961680 & -1.5736806 & $-0.35\pm0.43$ & $-0.57\pm0.68$ & $25.68\pm0.33$ & \\ 
8.3 & 39.983911 & -1.5733471 & $-0.15\pm1.50$ & $ 0.00\pm1.98$ & $26.98\pm1.34$ & \\ 
\hline
9.1 & 39.962215 & -1.5778958 & $ 0.40\pm0.74$ & $-0.22\pm0.99$ & $27.41\pm0.59$ & [1.54$\pm$0.06] \\
9.2 & 39.969287 & -1.5762569 & $ 0.28\pm1.34$ & $ 0.44\pm1.27$ & $27.47\pm1.03$ & \\ 
9.3 & 39.981833 & -1.5765536 & $ 0.04\pm0.49$ & $-0.02\pm0.50$ & $26.37\pm0.34$ & \\ 
\hline
10.1 & 39.968189 & -1.5714444 & $ 0.10\pm0.47$ & $ 0.50\pm0.41$ & $26.05\pm0.35$ & \\ 
\end{tabular}
\end{table*}

An 88 ksec {\it Chandra} ACIS observation was obtained in October 1999 as part of the program PID 525 (PI: Garmire).  
We used the adaptively smoothing algorithm {\tt csmooth} in the CIAO package \citep{asmooth} on the events file with a maximum scale of 15\arcsec. The astrometry of the X-ray data was checked using two bright isolated X-ray point sources located in the ACS image, and is better 
than 0.5\arcsec.

\section{Strong-lensing analysis}
\label{sl}
We have used Lenstool\footnote{Publically available, see {\tt http://www.oamp.fr/cosmology/lenstool} to 
download the latest version}\citep{Kneib93,Jullo} to perform a mass reconstruction of the cluster, assuming a parametric  
model for the distribution of dark matter. This model was constrained using the location of all multiple images identified.
This technique builds on the work done in \citet{Kneib96}, \citet{Smith05} and \citet{Richard09b}, 
by describing the cluster mass distribution with a superposition of analytic mass components to account for both cluster- and galaxy-scale 
mass. For each component we use a dual pseudo-isothermal elliptical mass distribution (dPIE, also known as a truncated PIEMD, \citealt{Ardis}). The dPIE profile is characterized by  its central position ($\alpha$,$\delta$), position angle ($\theta$), ellipticity ($e$), fiducial velocity dispersion $\sigma_0$ and two characteristic radii: a core radius $r_{\rm core}$ and a cut-off radius $r_{\rm cut}$ . 
Galaxy-scale mass components are added at the location of galaxies color-selected from the cluster red sequence. 
 In order to limit the number of free parameters, the velocity dispersion, cut-off radius and core radius is scaled to the galaxy luminosity $L$, relative to the luminosity of a $L^*$ galaxy \citep[see][for details]{Jullo}.
We fix  r$_{\rm core}^*$=0.15 kpc  and r$_{\rm cut}^*$=45 kpc to prevent degeneracies between the different parameters (see also \citealt{Limousin07} and Richard et al.\ 2009b for a discussion of these parameters). 
 We kept $\sigma_0^*$ as a free parameter. Finally, the unknown redshifts of systems 3 to 9 are kept as free parameters. The  faint surface brightness of the northern radial arc (System 10) and the absence of counter-images do not allow us to include it as a constraint.

The model is optimized with the Bayesian Markov chain Monte-Carlo (hereafter MCMC) sampler,
described in detail in \citet{Jullo}. The mass distribution is optimized by minimising the distance between the 
positions of the multiple images unlensed back to the source plane. We used the image-plane root mean squared (RMS) 
distance $\sigma_i$ of the images predicted by the model to the observed positions as an accuracy estimator 
of the model \citep{Limousin07}. 

A model including a single cluster-scale component, such as the one used in the early analysis by \citet{Bergmann}, largely fails in reproducing the multiple constraints ($\sigma_i\sim$3.5\arcsec). This result was
already pointed out by \citet{Kneib93b}, who showed that the locations and curvature radii of the arcs and multiple images (in particular the 
systems 1 and 2) can only be reproduced using a bimodal mass distribution. Indeed, we find that a model with 2 cluster clumps $DM1$ and $DM2$ fixed at the locations of the 
two brightest cluster galaxies gives a much better $\sigma_i$ (2.2\arcsec). Thanks to the 
larger number of identified multiple images systems, we can allow the 
 centers ($\alpha_1,\delta_1$) and ($\alpha_2,\delta_2$) of each clump to vary.

In addition, the shape and location of the 5 images previously identified in the giant arc (System 2) are strongly influenced by the nearby 
cluster galaxies. This is particularly true for the galaxy GAL1 (Fig \ref{mulfig}) which distorts the circular shape of the arc, and the BCG at the south hand side of the cluster. To better reproduce the giant arc, we choose to model 
these two galaxies with two independent dPIE potentials. In total, the new model contains 24 free parameters and we have 
40 constraints from the multiple images.
The resulting $\sigma_i$ of this model is $\sim 1.76$\arcsec, which is good compared  
to other similar works \citep{Limousin07,Richard09}. We adopt this model as our best-fit model for the rest of this letter.
The parameters of this model as well as their 1 $\sigma$ errors are given in Table \ref{bestpar}.
The predicted redshifts for systems 3 to 9 are presented in Table \ref{mult}. 
We note that  the large majority of them are about $z\sim1$. Such a redshift overdensity is often seen when looking at lensed fields.


%
\begin{table*}
\caption{\label{bestpar}Best fit parameters of the mass models. Values in brackets are not optimized.}
\begin{tabular}{llllllll}
Comp. & $x$ & $y$  & $e$ & $\theta$ & $\sigma_0$ & $r_{\rm core}$ & $r_{\rm cut}$ \\
                      & $[\arcsec]$ & $[\arcsec]$ &    & $[deg]$ &  $[km/s]$ & kpc & kpc \\
\hline
DM1 & $1.2\pm0.2$ & $-0.6\pm0.6$ & $0.26^{+0.10}_{-0.06}$ &  $-11\pm{4.0}$ & $596\pm30$ & $30.2^{+3.3}_{-0.2}$ & [800] \\
DM2 &  $1.5\pm0.2$& $27.3\pm1.0$ &  $0.31^{+0.02}_{-0.03}$ & $-6.1\pm0.6$ & $1316\pm35$ & 
$123^{+8}_{-4}$ & [800] \\
BCG & [0.0] & [0.0] & [0.30] & $[-81.9]$ & $194\pm30$ & [0.14]& $43\pm5$\\
GAL1 & [7.9] & [$-9.8$] & [0.26] & $[25.7]$ &  $118\pm13$ & [0.06] & $20\pm6$ \\
L$^{*}$ &      &        &          &            &   $193\pm9$ & [0.15] & [45] \\
\end{tabular}
\end{table*}

Although the value of $\sigma_0$ for $DM2$ is a lot higher than for $DM1$, 
$r_{core}$ for $DM2$ is also much larger, and therefore the 
 total masses of each cluster-scale component are similar. 
 Thanks to the proximity of GAL1 to images 2.3, 2.4 and 2.5 (see  Fig.~\ref{mulfig}), 
we obtain tight constraints on its parameters $\sigma_1$ and $r_{{\rm cut},1}$ (Table \ref{bestpar}). 
Interestingly, we find very little difference between the constrained values ($\sigma_1=118$ km/s, $r_{{\rm cut},1}$=20 kpc) and the values it would have had if scaled according to its luminosity 0.17 $L^*$ ($\sigma_1^{'}=123$ km/s, $r_{\rm cut}^{'}$=18 kpc).

\section{Results}
\label{results}

We use the best-fit model to unlens 
the giant arc to the source plane. Fig. \ref{mulfig}
shows that the source morphology resembles a local spiral galaxy with a red bulge, blue spiral arms containing 
individual knots of star formation. We also overlay the location of the cluster-scale and GAL1 caustic lines  at the giant arc redshift $z=0.725$. They part the source into 3 regions of different multiplicity. 
The eastern part of the source lies outside of both caustics and is singly imaged, the western part inside the cluster caustic 
is triply imaged, while the central red bulge inside both caustics is a system of 5 images, as we identified earlier from the HST image. The reconstruction of each part of the giant arc agrees with this source-plane morphology. 
Intrinsically, the source extends 10.0 / 2.5 kpc along its major/minor axis, and assuming a disk-like morphology it is observed with an inclination of 75 degrees. 
By comparing the overall fluxes in the image and the source plane, we derive a total magnification of $32\pm4$ for the entire arc. The largest linear 
magnification is obtained for image 2.2, where the centre is resolved at $<50$ pc. 

\begin{figure}
\centerline{\mbox{\includegraphics[width=7cm,angle=0]{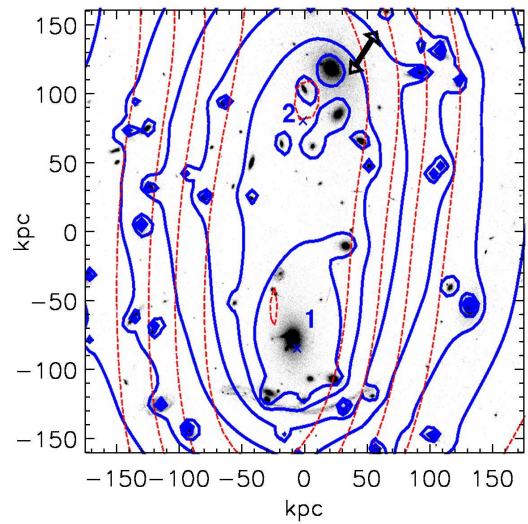}}}
\caption{\label{xray}Total mass (blue solid) and Xray (red dashed) lineary-spaced contours overlayed on top of the HST image. The centers of the cluster-scale 
clumps are marked with 1 and 2, and the orientation of the radial arc system 10 is indicated by a double arrow.}
\end{figure}

The reconstructed mass distribution (Fig. \ref{xray}) shows two main peaks located 
over the two brightest cluster galaxies. The center of $DM1$ is located within 1\arcsec of the BCG, while 
a significant offset (about 10\arcsec, or 50~kpc)
separates the center of $DM2$ from the second BCG to the north.  
This offset is well constrained by the presence of multiple systems on both northern and southern directions. Such a large offset between the dark matter component 
and the stellar light is not unusual, as a similar value of $\sim$9\arcsec was found in the case of Abell 1689 \citep{Limousin07}. 
We also observe that the orientation of the northern radial arc (System 10), which was \textit{not} explicitly included as a strong lensing constraint, is fully 
consistent with the location of $DM2$, rather than with the center of the northern BCG (Fig. \ref{xray}). 
Finally, we note that the model in which  the centers of $DM1$ and $DM2$ are fixed at the positions of the 
respective BCGs  (see Sect. \ref{sl}) requires a very large and unrealistic ellipticity for the secondary clump ($e\sim0.6-0.7$). 

The mass enclosed within a radius of 250 kpc from the barycenter of the mass distribution (located halfway between the centers of $DM1$ and $DM2$) is 3.8$\pm$0.2 10$^{14}$ M$_{\odot}$. 
We find a good agreement (within 3$\sigma$) when deriving the same mass using the NFW fit of the Subaru weak-lensing measurements from 
\citet{Broadhurst08a}: $M_{\rm WL}$=4.3 10$^{14}$ M$_{\odot}$. The statistical error on the enclosed mass is always better than 5\% within a radius of 
100\arcsec\ , which is a significant improvement compared to the model of \citet{Kneib93b} who found typical errors of $\sim15\%$. Then, we estimated the \textit{effective} Einstein radius $\theta_{\rm E}$, defined for a source at $z=2.0$, as the radius at which the averaged convergence $\overline{\kappa(<\theta_{\rm E})}=1$ (see \citealt{Broadhurst08b,Richard09}). By computing it from the barycenter of the mass distribution, we find $\theta_{\rm E}$=39\arcsec$\pm$2\arcsec. The enclosed mass within $\theta_{\rm E}$ is 2.82$\pm$0.15 10$^{14}$ M$_{\odot}$.  This large Einstein radius makes Abell 370 one of the best clusters to be used as a \textit{gravitational telescope} to search for very distant galaxies, similar to other clusters such as A1689 \citep{Broadhurst05}, A1703 \citep{Richard09} or MACS0717 \citep{Zitrin09c}.  Indeed, we outline in Fig. \ref{mulfig} the boundary of the region where
multiple images happen for a source at very high redshift ($z=6$), and show that it can assimilate to a circular region of $\sim50\arcsec$ radius.

In Fig. \ref{xray}, we overlay the X-ray luminosity contours on top of the  mass distribution contours, and observe 
a very good match. In particular, the contours share a similar elliptical shape and the  
same orientation. We use the {\tt ellipse} IRAF task to quantify the typical ellipticity $e$ and position angle $\theta$ of these contours, and measure ($e=0.60,\theta=-6^\circ$) 
and ($e=0.62,\theta=-5^\circ$) for the dark matter and X-ray maps, respectively, confirming the very good morphological agreement. Like the mass distribution, the X-ray luminosity map shows evidence for two prominent maxima, located near the dark matter clumps, with an offset smaller than 10\arcsec. 

Overall, the galaxy distribution, the mass distribution and the X-ray luminosity map each present two consistent 
peaks, and 
suggest that we are witnessing a massive cluster during a phase of merging. Only two other clusters have shown such a bimodality with two  peaks of dark matter and X-ray: the 
\textit{bullet cluster} \citep{Bradac07,Bradac09}, and MACS0025 (also known as the \textit{baby bullet}, \citealt{macs0025}). The main difference between 
these two clusters and Abell 370 is that both show much larger offsets (of the order of 200-350 kpc) between the X-ray and dark matter centers on each peak. 
Therefore it seems that the smaller offsets in Abell 370 are due to a projection effect, or due to the fact that the cluster is seen in an earlier stage of the 
merging process, at a time when the merging process has not yet affected much the baryons in the ICM.

Looking at the dynamical information on Abell 370, \citet{a370vdisp} have measured the redshift distribution of cluster galaxies, and shown that the redshift 
distribution has two redshift peaks separated by $\sim 3000$ km/s in velocity, with each of the two brightest galaxies belonging to a different 
redshift group. This suggests that contrary to the case of the bullet cluster and MACS0025, the merging of the two cluster components has a large 
projected velocity along the line of sight, which partly  explains the smaller offsets observed between the X-ray and dark matter peaks. 

\section*{Acknowledgments}
We are very grateful to the astronauts who made the service mission SM4 a scientific success, and to K.Noll, R.Lucas and their team who have conducted this spectacular ERO observation. We acknowledge useful discussions with Mark Swinbank.  JR acknowledges support from an EU Marie-Curie fellowship. 
JPK aknowledges support from the CNRS, the Agence Nationale de la Recherche grantÊANR-06-BLAN-0067, and the French-Israelian collaboration projectÊ07-AST-F9. ML acknowledges the CNES and CNRS for  their support. The Dark Cosmology Centre is funded by the Danish National Research  Foundation. Results are based on observations made with the NASA/ESA Hubble Space Telescope.

\bibliography{references}

\label{lastpage}

\end{document}